\documentclass[%
superscriptaddress,
preprint,
amsmath,amssymb,
aps,
]{revtex4-1}

\usepackage{lineno,hyperref}
\usepackage{amsmath}
\usepackage{amssymb}
\begin{document}


\title{Probability representation of quantum channels}


\author{Ashot Avanesov}
\email[]{avanesov@phystech.edu}

\affiliation{Department of General and Applied Physics, Moscow Institute of Physics and Technology (State University) Institutskii per. 9, Dolgoprudnyi, Moscow Region 141700, Russia}

\affiliation{Lebedev Physical Institute, Russian Academy of Sciences\\ Leninskii Prospect 53, Moscow 119991, Russia}

\noaffiliation

\author{Vladimir I. Man'ko}
\email[]{manko@lebedev.ru}
\affiliation{Department of General and Applied Physics, Moscow Institute of Physics and Technology (State University) Institutskii per. 9, Dolgoprudnyi, Moscow Region 141700, Russia}


\affiliation{Lebedev Physical Institute, Russian Academy of Sciences\\ Leninskii Prospect 53, Moscow 119991, Russia}

\affiliation{Tomsk State University, Department of Physics, Lenin Avenue 36, Tomsk 634050, Russia}

\date{\today}

\begin{abstract}
Using the known possibility to associate   the completely positive maps with density matrices
and recent results on expressing the density matrices with sets of classical probability distributions of dichotomic random variables we construct the probability representation of the completely positive maps.
In this representation, any completely positive map of qudit state density matrix is identified with the set of classical coin probability distributions.
Examples of the maps of qubit states are studied in detail. The evolution equation of quantum states is written in the form of the classical-like kinetic equation for probability distributions identified with qudit state.
\end{abstract}


\maketitle

\section{Introduction}

It is known that any quantum operation of quantum states defined in a $d$-dimensional Hilbert space $\mathcal{H}_d$ corresponds to a nonnegative hermitian operator acting on the tensor product of Hilbert spaces $\mathcal{H}_d\otimes\mathcal{H}_d$ \cite{bengtsson_geometry_2007}. Channel-state duality or Choi-Jamio{\l}kowski isomorphism \cite{choi_completely_1975,jamiolkowski_linear_1972} has a fundamental physical meaning and vastly applicated in the field of quantum information theory.

Since the origin of quantum mechanics, the pure states of quantum systems are described by wave function or wave vector that is an element of a Hilbert space \cite{schrodinger_undulatory_1926,dirac_principles_1982}. The quantum state also can be determined by the density matrix \cite{landau_dampfungsproblem_1927,neumann_wahrscheinlichkeitstheoretischer_1927} that is a nonnegative hermitian operator acting on the Hibert space. Moreover, it is impossible to present the state of a quantum system in the presence of thermal fluctuation in the form of a wave function. Naturally, the density matrix describes these states called mixed states.

Throughout the whole period of the development of quantum mechanics and its applications, the alternative approaches to describe quantum states were being suggested. Some of them initially were developed in attempts to construct the formalism of quantum mechanics that is similar to classical statistical mechanics one \cite{wigner_quantum_1932,husimi_formal_1940,kano_new_1965,glauber_coherent_1963,sudarshan_equivalence_1963}. Authors of these works identified quantum states with different kinds of the quasiprobability distributions which are functions of position and momentum.

Also, there were attemts to construct the representation of quantum mechanics based only on fair probability distribution \cite{mielnik_geometry_1968,wootters_wigner-function_1987,khrennikov_classical_2019,khrennikov_quantum_2019}. Some approaches proposed the usage of non-Kolmogorov probability theories \cite{khrennikov_non-kolmogorov_1997}.

In \cite{mancini_symplectic_1996} it was introduced the tomographic probability representation of states of the quantum system with continuous variables, where the quantum state is associated with fair probability distributions. This representation relates to the notion of optical tomogram function which was introduced in \cite{bertrand_tomographic_1987,vogel_determination_1989} where authors proposed to measure tomogram probabilities in purpose to reconstruct the Wigner function of a photon quantum state. The approach of \cite{mancini_symplectic_1996} was expanded for the case of systems with discrete variables and notion of spin-tomogram was introduced \cite{dodonov_positive_1997,manko_spin_1997}. For further information on the tomographic probability representation of quantum mechanics, we refer to review \cite{ibort_introduction_2009}.

The evolution of closed quantum systems is described by the Schr\"{o}dinger equation \cite{schrodinger_quantisierung_1926}. Hence, the state of the closed system undergoes the unitary transformation. It is also important to consider the problem of the evolution of open quantum systems. The general nonunitary transformation of the quantum state is determined by completely positive map of the density matrix and can be presented in the form \cite{stinespring_positive_1955}
\begin{equation}
\label{quantum channel}
\hat\rho \rightarrow\sum_{k}\hat A_k\hat\rho\hat A_k^\dagger,
\end{equation}
where $\hat\rho$ is the initial density matrix of the state and $\hat A$ is an arbitrary matrix called Kraus operator \cite{kraus_general_1971}. The main aspects of the map (\ref{quantum channel}) are widely discussed in the literature, and it is also known as quantum channel, e. g. see \cite{nielsen_quantum_2010}. The notion of dynamical maps were introduced in \cite{sudarshan_stochastic_1961,jordan_dynamical_1961}.  The evolution equation of open quantum systems was derived in \cite{gorini_completely_1976,lindblad_generators_1976}.

In the case of qudit systems, the state can be reconstructed from the finite number of values of the tomographic function. For instance, we need only three real parameters to determine the state of qubit \cite{bloch_nuclear_1946}. Hence, we can use three probabilities to describe the qubit state. 
Recently, quantum suprematism or probability representation of quantum mechanics of discrete variables systems was suggested \cite{chernega_quantum_2018,chernega_probability_2017,lopez-saldivar_geometry_2018,chernega_triangle_2017-1}. Initially, it was proposed for the case of qubit systems. Here, the state associated with three probability distributions based on tomographic probabilities, the quantum observable is treated as the set of classical-like variables, though the dependence between statistics of quantum observable and its classical-like variables is not straightforward \cite{chernega_correlations_2019}. Generalization to the description of qudit states in terms of the probabilities also was proposed \cite{chernega_triangle_2017}.

The goal of our approach is to present the formulation of quantum states and the state evolution equations
in terms of probability distributions and stochastic (classical-like) kinetic equations.
In the present paper, we are aimed to utilize the channel-state duality in the framework of the developed probability representation and thus to express the quantum operation as the set of probability distributions. Then, it is also possible to derive the stochastic equation for these probabilities that correspond to the evolution equation of quantum systems. In the present paper, we consider the unitary evolution of the qubit system.

\section{Qudit states in probability representation}
\label{sec:probability representation}

Let us give a brief review of the main aspects of probability representation of the states of the finite-level quantum (qudit) systems. This approach is directly derived from the notion of spin tomographic function that is introduced for spin systems. In the case of spin $j$ it has the expression
$
w(m,\ \vec n) = \langle m|\hat U(\vec n)\hat\rho\hat U^\dagger(\vec n)|m\rangle
$,
where $m=-j,\ -j+1,...,\ j$ is spin projection onto direction determined by the vector $\vec n$, $|m\rangle$ is eigenvector in computational basis (z-basis). The unitary matrix $\hat U(\vec n)$ is chosen by the way that $\hat U^\dagger(\vec n)|m\rangle$ is eigenvector of operator of spin projection onto direction $\vec n$, that is expressed in the form $n_x\hat\sigma_x+n_y\hat\sigma_y+n_z\hat\sigma_z$. Here $\hat\sigma_x$, $\hat\sigma_y$, $\hat\sigma_z$ are Pauli matrices. Thus, the value $w(m,\ \vec n)$ is the probability of the outcome $m$ the measurement of spin projection onto direction of vector $\vec n$.

The knowledge of tomographic function allows us to reconstruct the density matrix of the state of the quantum system. Moreover, we need only a finite number of its values. To be precisely in the case of the $n$-level quantum system its density matrices determined by $n^2-1$ real parameters.  Finally, we can present these parameters as the tomographic probabilities.

For example, in the case of qubit systems, we need only three tomographic probabilities to describe the state $
p_1 = \langle +|\hat U_x\hat\rho\hat U^\dagger_x|+\rangle,\quad p_2 = \langle +|\hat U_y\hat\rho\hat U^\dagger_y|+\rangle,\quad p_3 = \langle +|\hat\rho|+\rangle
$,
where $\hat U_x = \frac{1}{\sqrt2}\begin{bmatrix}
1 & 1\\1 & -1
\end{bmatrix}$, $\hat U_y = \frac{1}{\sqrt2}\begin{bmatrix}
1 & i\\ i & 1
\end{bmatrix}$ and $|+\rangle = \begin{bmatrix}
1\\0
\end{bmatrix}$. In other words $p_1 = w(+,\ x)$, $p_2=w(+,\ y)$, and $p_3=w(+,\ z)$. Then, the density matrix of the qubit state can be presented in the form

\begin{equation}
\hat\rho = \begin{bmatrix}
p_1 & \left(p_2-\frac12\right)-i\left(p_3-\frac12\right)\\
\left(p_2-\frac12\right)+i\left(p_3-\frac12\right)& 1-p_1
\end{bmatrix} .
\end{equation}
These probability parameters form three dichotomic probability distributions. However, they must satisfy the restriction $
\sum_{i=1}^3\left(p_i-\frac12\right)^2 \leq \frac14
$.

In essence, in the probability representation of the qudit systems states are described by the finite set of probability distributions. In the case of qubit systems, we use the three distributions corresponding to the measurements of spin projection onto three perpendicular directions $x$, $y$ and $z$.

In the case of ququart systems, the probability parametrization of the density matrix can be presented in the form

\begin{widetext}
	
	\begin{equation}
	\label{ququart parametrization}
	\hat\rho = \begin{bmatrix}
	p_1 + p_2 + p_3 - 2 & \left(p_4-\frac12\right) - i\left(p_5-\frac12\right) & \left(p_6 - \frac12\right) - i\left(p_7-\frac12\right) & \left(p_8-\frac12\right)-i\left(p_9-\frac12\right)\\
	\left(p_4-\frac12\right) + i\left(p_5-\frac12\right) & 1-p_1 & \left(p_{10}-\frac12\right) - i\left(p_{11}-\frac12\right) & \left(p_{12}-\frac12\right) - i\left(p_{13}-\frac12\right) \\
	\left(p_6-\frac12\right) + i\left(p_7-\frac12\right) & \left(p_{10}-\frac12\right) + i\left(p_{11}-\frac12\right) &
	1-p_2 & \left(p_{14}-\frac12\right)-i\left(p_{15}-\frac12\right)\\
	\left(p_8-\frac12\right)+i\left(p_9-\frac12\right) & \left(p_{12}-\frac12\right)+i\left(p_{13}-\frac12\right) & \left(p_{14}-\frac12\right)+i\left(p_{15}-\frac12\right) & 1-p_3
	\end{bmatrix}
	\end{equation}
	
\end{widetext}

Thus, we can determine the state of the quantum system as the set of $12$ dichotomic probability distributions and one of the size $4$. In other words, instead of density matrix we describe the ququart state as a set of probability distributions $
\Xi^{(4)} = \left\{\vec P,\ \ \vec P_i,\ i=4,...,\ 15 \right\},
$
where

\begin{equation}
\vec P = \begin{bmatrix} p_1+p_2+p_3\\ 1-p_1\\1-p_2\\1-p_3 \end{bmatrix},\quad \vec P_i = \begin{bmatrix}
p_i\\1-p_i
\end{bmatrix}.
\end{equation}
As we see a bit further this representaion of ququart states will be useful in the construction of probability representation of completely positive maps of qubit systems.

\section{Completely positive maps and Choi-Jamio{\l}kowski isomorphism}
\label{sec:completely positive maps}

The general linear transformation $F$ of the quantum system states can be presented in the form

\begin{equation}
F[\hat\rho]_{ij} = \sum_{i_0=1}^d\sum_{j_0=1}^dD_{ii_0,\ jj_0}\rho_{i_0j_0},
\end{equation}
where $\rho_{i_0j_0}$ denote the elements of the density matrix $\hat\rho$.

If operation $F$ preserves the hermiticity and nonnegativity of eigenvalues of the matrix transforming matrix $\hat\rho$, then the map $F$ is called positive. We also can add the requirements of trace-preserving, so the matrix $F[\hat\rho]$ would be the density one.

By using the aforementioned requirements of preserving hermiticity and trace we can conclude that $D_{ii_0,\ jj_0} = D^*_{jj_0,\ ii_0}$ and $\sum_{i=1}^d D_{ii_0,\ ij_0} = \delta_{i_0j_0}$.

Finally, let us recall the definition of completely positive maps. Here, in addition to the examined $d$-level quantum system $\mathcal{H}_A$, we introduce the $n$-level environment $\mathcal{H}_E$ and consider the maps of the whole system $\mathcal{H}_A\otimes\mathcal{H}_E$. In particular, we are interested in the maps of the form $I_n\otimes F$, where $I_n$ is the identical operator which acts in the space of the states of the system $\mathcal{H}_E$ and $F$ is the transformation of the states of the system $\mathcal{H}_A$.
If for every $n$ the map $I_n\otimes F$ is positive, then the map $F$ is a completely positive one.

Let us introduce the matrix of the form
\begin{equation}
\hat D = \sum_{k,l,i,j=1}^{d}D_{ki,\ lj}|k\rangle\langle l|\otimes|i\rangle\langle j|.
\end{equation}
The matrix $\hat D$ accordingly to \cite{sudarshan_stochastic_1961} is called dynamical. Due to the properties of positive maps, it is hermitian one that has the trace equal to $d$.

Finally, the Choi theorem tells us that the dynamical matrix of the completely positive map can only have nonnegative eigenvalues.

Thus, we briefly reviewed the main aspects of the description of completely positive maps. The corresponding to the transformation of $d$-level system dynamical matrix $\hat D$ is the Hermitian, positive-semidefinite matrix with the fixed trace (actually $\mathrm{Tr}\hat D =d$). Hence, the matrix $\frac{1}{d}\hat D$ satisfies all the requirements of the density matrix. The Choi-Jamio{\l}kowski isomorphism makes the correspondence between the completely positive maps of $d$-level systems and the density matrices of $d^2$-level systems.

\section{Probability representation of completely positive maps}
\label{sec:probability and completely posistive maps}

As it was shown, the set of $d^2-1$ probability parameters (that was contained in one vector $\vec{\mathcal{P}}$) determined the state of $d$-level quantum system. From these probabilities we can construct $N-1 = d^2 - d$ dichotomic distributions and one distribution of size that equal to $d$. We denote the set of these distributions as $\Xi^{(d)}$.

We remind that there is the isomorphism between the completely positive maps of the states of $d$-level systems and the states of $d^2$-level systems. Therefore, we can use the introduced probability parametrization of the normalized dynamical matrix $\frac{1}{d}\hat D$. By accomplishing this procedure, we come to probability representation of the transformation of quantum states.  

Let us demonstrate the possibility of probability description of completely positive maps on the example of qubit systems.

The dynamical matrix of the completely positive map of qubit state is hermitian and its trace is equal to $2$. We can utilize the parametrization of ququart state (\ref{ququart parametrization}). We also should note that the probability parameters that determine the considered map must satisfy the following relations $
p_1+p_3 = \frac32\quad
p_4 + p_{14} = 1\quad
p_5 + p_{15} = 1
$.

Hence, the completely positive map of qubit state is described by the vector $\vec{\mathcal{P}}$ of $15$ probability parameters or by the set of probability distributions $\Xi^{(4)}$. The vector $\vec{\mathcal{P}}$ can be obtained from the vectorized matrix $\hat D$. By vectorizing we mean the procedure when we take raws of the initial matrix of size $n\times n$ and put them in one raw accordingly to their position in the matrix, then by transposition operation obtain the vector of size $n^2$. In other words the product of vectorization of matrix $\hat D$ is vector $\vec D$ of the form

\begin{equation}
\vec D = \sum_{k,l,i,j=1}^{d}D_{ki,\ lj}|k\rangle|i\rangle|l\rangle|j\rangle .
\end{equation}
The probability vector $\vec{\mathcal{P}}$ is a linear transform of vectorized dynamical matrix

\begin{equation}
\vec{\mathcal{P}} = \hat A\cdot\vec{D} + \vec b
\end{equation}
where $\hat A$ is a matrix of size $15\times 16$. Nonzero elements of the matrix are $A_{1,\ 6}=A_{2,\ 11}=A_{3,\ 16}=-\frac12$, $A_{4,\ 2}=A_{4,\ 5}=A_{6,\ 3}=A_{6,\ 9}=A_{8,\ 4}=A_{8,\ 13}=A_{10,\ 7}=A_{10,\ 10}=A_{12,\ 8}=A_{12,\ 14}=A_{14,\ 12}=A_{14,\ 15}=\frac14$ and $A_{5,\ 2}=-A_{5,\ 5}=A_{7,\ 3}=-A_{7,\ 9}=A_{9,\ 4}=-A_{9,\ 13}=A_{11,\ 7}=-A_{11,\ 10}=A_{13,\ 8}=-A_{13,\ 14}=A_{15,\ 12}=-A_{15,\ 15}=\frac{i}{4}$. All elements of vector $\vec b$ except $b_1$, $b_2$ and $b_3$ are equal to $\frac14$. For the first three entries of vector $\vec b$ we have $b_1=b_2=b_3=\frac12$. Note, that we have the map $\mathbb{R}^{16}\rightarrow\mathbb{R}^{15}$, that is why it is possible to use other transformation matrix $\hat A$ and vector $\vec b$.

We are also able to reconstruct the dynamic matrix $\hat D$ from the given vector of probability parameters $\vec{\mathcal{P}}$. Here, it is also presented as the linear map

\begin{equation}
\vec D = \hat B\cdot\vec{\mathcal{P}} + \vec c
\end{equation}
Nonzero elements of matrix $\hat B$ are $B_{1,\ 1}=B_{1,\ 2}=B_{1,\ 3}=B_{2,\ 4}=B_{3,\ 6}=B_{4,\ 8}=B_{5,\ 4}=-B_{6,\ 1}=B_{7,\ 10}=B_{8,\ 12}=B_{9,\ 6}=B_{10,\ 10}=-B_{11,\ 2}=B_{12,\ 14}=B_{13,\ 8}=B_{14,\ 12}=B_{15,\ 14}=-B_{16,3}=2$ and $B_{2,\ 5}=B_{3,\ 7}=B_{4,\ 9}=-B_{5,\ 5}=B_{7,\ 11}=B_{8,\ 13}=-B_{9,\ 7}=-B_{10,\ 11}=B_{12,\ 15}=-B_{13,\ 9}=-B_{14,\ 13}=B_{15,\ 15}=-2i$. For elements of vector $\vec c$ we have $c_1=-4$, $c_6=c_{11}=c_{16}=2$, $c_2=c_3=c_4=c_7=c_8=c_{12}=-1+i$ and $c_5=c_9=c_{10}=c_{13}=c_{14}=c_{15}=-1-i$.

\section{Stochastic equations}
\label{sec:evolution}

The evolution process can be described by the completely positive map that depends on the time parameter. We have shown that it is possible to use the probability distributions to describe the transformations of quantum states. In order to develop our approach, we decide to find the equation for the vector of probability parameters $\vec{\mathcal{P}}$.
In the present paper, it is considered a unitary evolution. In the case the dynamical matrix can be expressed in the form $\hat D = \vec U\cdot\vec U^\dagger$, where $\vec U$ is vectorized unitary matrix $\hat U$ that obeys to Shr\"{o}dinger equation.
Then, the evolution equation for the dynamical matrix $\hat D$ takes the form

\begin{equation}
i\frac{d\hat D}{dt} = \left[\hat H\otimes\hat I_d,\ \hat D\right],
\end{equation}
where $[\hat A,\ \hat B]$ is commutator of operators $\hat A$ and $\hat B$ and $\hat H$ is Hamiltonian of the system. Thus, the dynamical matrix obeys the von Neumann equation for the $d^2$-level system described by the Hamiltonian $\hat H\otimes \hat I_d$. Thus, we can obtain the evolution equation for vector $\vec D$

\begin{equation}
i\frac{d\vec D}{dt} = \hat Q\cdot\vec D
\end{equation}
where $\hat Q = \hat H\otimes\hat I_8-\hat I_8\otimes\hat H$ .

Finally, we come to the description of the evolution process in terms of the probability representation.  The equation for the vector of probability parameters $\vec{\mathcal{P}}$ has the form

\begin{equation}
\label{kinetic equation}
i\frac{d\vec{\mathcal{P}}}{dt} = \hat A\hat Q\hat B\cdot\vec{\mathcal{P}} + \hat A\hat Q\cdot\vec{c}
\end{equation}

We should add the initial condition to the last equation so that the problem would have a unique solution. In the time moment, $t=0$ the transformation matrix $\hat M$ is the unity matrix. Therefore, the corresponding dynamic matrix $\hat D$ after normalization is the density matrix of maximally entangled state $\hat D^{(0)} = (|0\rangle|0\rangle + |1\rangle|1\rangle)(\langle 0|\langle 0| + \langle 1|\langle 1|)$.
Thus, we finally derive that all initial probability parameters are equal to $\frac12$ except the following ones $p_1^{(0)} = p_2^{(0)} = p_8^{(0)} = 1,\quad p_3^{(0)} = \frac12
$.

\section{Summary}
\label{sec:summary}

To conclude we point out the main results of the work. We demonstrated on an example of qudits that the quantum states can be identified
with probability distributions. A new aspect of this work is that we expressed the matrix elements of the evolution operator of the qudit system
with given Hamiltonian in terms of probabilities and the time evolution equation for the unitary evolution of the system
is presented in the form of classical-like kinetic equation (\ref{kinetic equation}) for these probability distributions.

The solutions of the kinetic equation are shown to provide the probability representation form of the completely positive quantum channels.

The approach developed in the paper can be extended to open system evolution and it will be done in a future publication.

\bibliography{refs}

\end{document}